\documentstyle[12pt]{article}
\global\parskip 6pt

\normalsize

\let\oldtheequation=\theequation
\def\doteqs#1{\setcounter{equation}{0}
            \def\theequation{{#1}.\oldtheequation}}

\newcounter{sxn}
\def\sx#1{\addtocounter{sxn}{1} \bigskip\medskip \goodbreak
\noindent{\large\bf 
{\thesxn.~~#1}} \nobreak \medskip}
\def\sxn#1{\sx{#1} \doteqs{\thesxn}}

\newcounter{axn}

\def\br{\begin{thebibliography}{abc}}
\def\er{\end{thebibliography}}

\def\be{\begin{equation}}
\def\ee{\end{equation}}
\def\bea{\begin{eqnarray}}
\def\eea{\end{eqnarray}}

\begin{document}
\begin{flushright}
\hfill{SINP-TNP/05-07}\\
\end{flushright}
\vspace*{1cm}
\thispagestyle{empty}
\centerline{\large\bf Asymptotic Quasinormal Modes of $d$ Dimensional Schwarzschild}
\centerline{\large\bf Black Hole with Gauss-Bonnet Correction}
\bigskip
\begin{center}
Sayan K. Chakrabarti\footnote{Email: sayan@theory.saha.ernet.in} 
and Kumar S. Gupta\footnote{Email: kumars.gupta@saha.ac.in (Corresponding
author) }\\ 
\vspace*{.5cm}
{\em Theory Division\\
Saha Institute of Nuclear Physics\\
1/AF Bidhannagar, Calcutta - 700064, India}\\
\vspace*{.5cm}
\end{center}
\vskip.5cm

\begin{abstract}
  
We obtain an analytic expression for the highly damped asymptotic
quasinormal mode frequencies of the $d \geq 5$-dimensional Schwarzschild black hole modified by the Gauss-Bonnet term, which appears in string derived
models of gravity. The analytic expression is obtained under the string inspired 
assumption that there exists a mimimum length scale in the system and in the limit when 
the coupling in front of the Gauss-Bonnet term in the action is small.
Although there are several similarities of this geometry with that
of the Schwarzschild black hole, the asymptotic quasinormal mode
frequencies are quite different. In particular, the real part of the
asymptotic quasinormal frequencies for this class of single horizon black
holes in not proportional to log(3).

\end{abstract}
\vspace*{.3cm}
\begin{center}
October 2005
\end{center}
\vspace*{1.0cm}
PACS : 04.70.-s \\
\newpage

\sxn{Introduction}

Quasinormal modes associated with the perturbations of a black hole
metric in classical general relativity  have been found to be a useful probe 
of the underlying space-time geometry \cite{rg,vish}.  The
quasinormal ringing frequencies carry unique information about black
hole parameters and they are expected to be detected in the future
gravitational wave detectors \cite{kk} and possibly even in the Large
Hadron Collider \cite{alex}. For asymptotically flat
space-times, these metric perturbations are solutions of the
corresponding wave equation with complex frequencies which are purely
ingoing at the horizon and purely outgoing at infinity \cite{chandra}.
For most geometries, the wave equation is not exactly solvable and
various schemes have been used in the literature to obtain approximate
analytical expression for the quasinormal modes. In particular, Motl
and Neitzke \cite {motl} have proposed a geometric method for
calculating the highly damped quasinormal modes in asymptotically flat
geometries. Their method involves the extension of the wave equation
beyond the physical region between the horizon and infinity by
analytically continuing the radial variable $r$ to the whole complex
plane. The asymptotic quasinormal modes are then obtained using
suitable monodromy relations in the complex plane that encode the
appropriate boundary conditions. This monodromy approach has
subsequently been applied to more general geometries
\cite{das,cardoso,natario,daghigh} and has also been extended to the
study of ``non-quasinormal modes" of various black holes as well
\cite{danny,ksg1}. 

In spite of their classical origin, it has recently been proposed that
the quasinormal modes might provide a glimpse into the quantum nature
of black holes. One such proposal by Hod \cite{hod} is associated with
the idea that black holes have a discrete area spectrum
\cite{7475ofsch}, with the area being quantized in integer multiples
of $4~{\rm log}(k)$, where $k > 1$ is an undetermined positive integer.
Hod's conjecture is based on the observation that the real part of the
asymptotic highly damped quasinormal modes of the Schwarzschild black
hole is independent of space-time dimensions as well as the nature of
the metric perturbation and proportional to ${\rm log}(3)$. This
universality strongly suggests that the real part of the highly damped
asymptotic quasinormal frequency of the Schwarzschild black hole is a
characteristic feature of the black hole itself. This observation
together with the area quantization law and first law of black hole
mechanics immediately leads to the conclusion that $k=3$. However, for
geometries other than Schwarzschild, the validity of Hod's conjecture
is debatable \cite{setare,das,natario}. In a related work, it was
shown that the real part of the asymptotic quasinormal mode of the
Schwarzschild black hole can be used to determine the Immirzi
parameter appearing in loop quantum gravity \cite{dryer}. Independent
of these connections, it has also been shown that the quasinormal
modes for various background geometries appear naturally in the
description of the corresponding dual CFT's living on the black hole
horizons \cite{adscft}.  These results have mostly been obtained for
geometries which are solutions of the equation of motion arising from
the classical Einstein-Hilbert action.

In this Paper we shall analyze the asymptotic quasinormal modes of the
$d$ dimensional Schwarzschild black holes in presence of a 
Gauss-Bonnet correction term\cite{gb1,gb2}, which appear when the
Einstein-Hilbert action is generalized to include the leading order
higher curvature terms arising from the low energy limit of string
theories \cite{zwei}. There has been a renewed interest in Gauss-Bonnet black
holes in the context of brane-world models \cite{fromross} and the
entropy and related thermodynamic properties of such black holes have
been discussed in recent literature \cite{ross}. The quasinormal
frequencies of certain low lying modes of the Gauss-Bonnet black hole
have also been estimated in the WKB approximation \cite{kon}. Our
interest is in the other end of the quasinormal frequency spectrum,
namely in the highly damped asymptotic regime. In addition, we want to study 
the properties of asymptotic quasinormal modes when the geometry is minimally different from the Schwarzschild background. In order to make this precise, we shall assume that there exists a fundamental minimum length scale in the system, in terms of which the smallness of the Gauss-Bonnet coupling would be specified. This assumption is justified by the fact that the Gauss-Bonnet term has its natural origin within the frameworks of string theory, which in turn postulates the existence of a fundamental minimum length scale in nature. In our analysis we shall use this string inspired assumption although the details of such a length scale would not be important. In the limit of a small Gauss-Bonnet coupling, the resulting classical geometry is very similar to Schwarzschild case. We shall however show that there are important
differences in the asymptotic form of the quasinormal modes. In
particular, the real part of the asymptotic quasinormal mode is
dimension dependent and is not proportional to ${\rm log}3$. Our result therefore encodes the effect of a small Gauss-Bonnet term on the asymptotic quasinormal modes of the $d$ dimensional Schwarzschild black hole.

This Paper is organized as follows. In Section 2 we shall briefly review the
Gauss-Bonnet term and discuss the modification of the $d$ dimensional
Schwarzschild metric in presence of a weakly coupled Gauss-Bonnet term in
the action. In Section 3 we shall apply the monodromy method to obtain the
asymptotic quasinormal modes of this system. Section 4 concludes the paper
with some discussions of our result and an outlook.

\sxn{The $d$-dimensional Schwarzschild metric with the Gauss-Bonnet 
term in the weak coupling limit}

In this Section we shall discuss some properties of the $d$-dimensional
Schwarzschild metric due to a Gauss-Bonnet term in the limit of small
Gauss-Bonnet coupling. In space-time dimensions $d \geq 5$, the 
Einstein-Hilbert action in presence of the Gauss-Bonnet term has the form
\be 
I=\frac{1}{16\pi } \left [ \int d^dx \sqrt{-g} R +
  \frac{\alpha}{(d-3)(d-4)} \int d^dx
  \sqrt{-g}(R_{abcd}R^{abcd}-4R_{cd}R^{cd}+R^2) \right ],\label{alpha}
 \ee 
 where we have set the Newton's constant $G_d$ in $d$ space-time
 dimensions and the velocity of light $c$ equal to one. The parameter
 $\alpha$ in Eqn. (\ref{alpha})is the Gauss-Bonnet coupling. We shall
 consider only positive $\alpha $, which is consistent with the string
 expansion \cite{gb1}.

The field equations can be written as:
\be 
\delta I/\delta g_{\mu\nu} = -G_{\mu\nu} + \alpha T_{\mu\nu} =0
\label{eqn} 
\ee 
where 
\be
T_{\mu\nu}=RR_{\mu\nu} - R_{\mu\alpha\beta\gamma}R_{\nu}^{~~\alpha\beta\gamma}
- 2R_{\alpha\beta}R^{\alpha~~\beta}_{~\mu~~\nu} -
2R_{\mu\alpha}R^{\alpha}_{~\nu} -
\frac{1}{4}(R_{\alpha\beta\gamma\delta}R^{\alpha\beta\gamma\delta} -
4R_{\alpha\beta}R^{\alpha\beta} + R^2).
\ee 
When $\alpha = 0$, the solution of the equations of motion is given by
the Schwarzschild metric. For small values of $\alpha$, the second
term in Eqn.\ (\ref{eqn}) would provide corrections to the Schwarzschild
geometry.

The metric for the spherically symmetric, asymptotically flat black
hole solution of mass $M$ arising from the action in Eqn.
(\ref{alpha}) is given by
\bea
ds^2 &=& - f(r) dt^2 + f^{-1} dr^2 + r^2 d \Omega^2_{d-2},\label{metric}\\
f(r)&=& 1+\frac{r^2}{2\alpha}-\frac{r^2}{2\alpha}
\sqrt{1+\frac{8\alpha M}{r^{d-1}}}, 
\eea 
where $r$ is the radial variable in $d$ space-time dimensions and
$\Omega^2_{d-2}$ is the metric on the $(d-2)$ dimensional sphere. 
Eqns. (2.4)-(2.5) give an exact solution to the equations of motion arising
from the action $I$ in (2.1). However, as mentioned before, we are interested 
in small deviations from the Schwarzschild geometry, which is obtained for a
small value of $\alpha$. Thus, in the limit $\alpha \rightarrow 0$, the 
function $f(r)$ in the metric (2.4) is given by
\be 
f(r)=1-\frac{2M}{r^{d-3}}+\frac{4\alpha M^2}{r^{2d-4}}\label{schw} 
\ee
The series expansion of Eqn. (2.5) can be done if $\frac{8\alpha
  M}{r^{d-1}}<<1$, which naturally leads to the question that is such an 
expansion valid for our analysis. In order to address this issue, note that the 
Gauss-Bonnet term has its natural origin in the framework of string theory 
which postulates the existence of a fundamental minimum length scale $l$, 
which is finite but can be arbitrarily small. 
The physics of any distance below this scale simply cannot be described in this 
context. In this scenario, if we choose $\alpha$ such that it satisfies the 
relation $\frac{8\alpha M}{l^{d-1}}<<1$, then the relation $\frac{8\alpha
  M}{r^{d-1}}<<1$ is satisfied for all $r > l$, i.e. for all distances that can be physically probed in this context. This necessarily restricts $\alpha$ to a very small range which is precisely what we require to obtain a small deviation from the 
Schwarzschild geometry. Thus, under this string inspired assumption of the 
existence of a minimum length scale $l$, Eqn. (2.6) indeed is valid as 
$\alpha \rightarrow 0$. In the series expansion of Eqn. (2.5) in powers of
$\alpha$, the last term in Eqn.\ (\ref{schw}) is physically the most
relevant term that captures the effect of a small Guass-Bonnet
correction to the Schwarzschild geometry.  Hence, in the rest of this
paper, we shall use the metric in Eqn.\ (\ref{metric}), with the
$f(r)$ given by Eqn.\ (\ref{schw}) to analyze the asymptotic
quasinormal modes of the $d$ dimensional Schwarzschild black hole in
presence of a small Gauss-Bonnet correction.

The black hole described above has a singularity at the origin
shielded by a single event horizon \cite{gb1,gb2}. In our
approximation, the horizon at $r = r_h$ is determined by the real
positive solution of the equation
\be
r_h^{2d-4} - 2M r^{d-1} + 4 \alpha M^2 = 0
\ee
To the leading order in $\alpha$, $r_h = r_s + \epsilon$, where $r_s$
is the location of the real horizon of the $d$-dimensional
Schwarzschild solution and the correction term $\epsilon \propto -
\alpha$. The exact form of the correction term is not important for us
and its negative sign is in agreement with the fact that in
Gauss-Bonnet black hole the horizon is always located at a magnitude
less than that in the case when $\alpha = 0$ \cite{gb2}. In our
approximation, apart from a single real horizon at $r_h$ given above,
there would be in general $(2d-5)$ other fictitious horizons with
complex or negative values of $r$.

\sxn{Asymptotic Quasinormal Modes}

In this Section we shall evaluate the asymptotic quasinormal mode
frequencies $w$ for the metric given by Eqn.\ (\ref{metric}) with the
form of the $f(r)$ given by Eqn.\ (\ref{schw}) in the limit when ${\rm
  Im}~(w) >>{\rm Re}~(w)$. In the absence of any numerical
calculations, at this stage it is an assumption that such modes indeed
exist. However, for small values of $\alpha$ this is a reasonable
assumption and our final result will also be shown to be consistent
with it. We shall consider the quasinormal modes in all space-time
dimensions $ d \geq 5$ except for $d=6$, as the metric perturbations
for the Gauss-Bonnet black hole in $d=6$ are known to be unstable
\cite{egb}. Below we shall closely follow the monodromy method of ref.
\cite{motl} and \cite{natario}.

The asymptotic quasinormal modes for the pure $d$ dimensional 
Schwarzschild black hole are found by solving a Schr\"{o}dinger like
equation with the Ishibashi-Kodama master potential \cite{ishi}. In presence
of the Gauss-Bonnet term, the tensorial perturbations describing the
quasinormal modes still follow a 
Schr\"{o}dinger like equation, but with a different potential \cite{egb}. In
this case, the Schr\"{o}dinger like equation is given by
\be              
\left [-\frac{d^2}{dx^2}+V[r(x)] \right] \Phi(x)=\omega^2
\Phi(x),\label{scho}  
\ee
where $x$ is the tortoise coordinate defined by $dx = \int
\frac{dr}{f(r)}$ and the potential $V$ is given by \cite{egb}
\be
V(r) = q(r)+
\left(f\frac{d}{dr}~ln~(K)\right)^2+f\frac{d}{dr}\left(f\frac{d}{dr}~ln~(K)\right)
\label{vr}
\ee
with $K(r)$ and $q(r)$ being given by 
\bea
K(r) & = & r^{\frac{d - 4}{2}}\sqrt{r^2~ + ~\frac{\alpha}{(d - 3)}\left[(d
    - 5)(1 - f(r))~ - ~rf^{\prime}\right]} \label{kr}\\
q(r) & = & \left(\frac{f(2 - \gamma)}{r^2}\right)\left(\frac{(1 - \alpha
    f^{\prime\prime}(r))r^2 + \alpha(d - 5)[(d - 6)(1 - f(r)) -
    2rf^{\prime}(r)]}{r^2 + \alpha(d - 4)[(d - 5)(1 - f(r)) -
    rf^{\prime}(r)]}\right) \label{qr},
\eea
where $\gamma=-l(l+d-3)+2$ and $l=2, 3, 4, \cdots$
The potential in Eqn.\ (\ref{vr}) reduces to the standard Ishibashi-Kodama
master potential for the Schwarzschild black hole when $\alpha = 0$.

If $\Phi (x)$ describes the quasinormal modes, then in terms of the
tortoise coordinate $x$ it must satisfy the boundary conditions
\bea
\Phi (x) &\sim& e^{i\omega x}\;\, {\mathrm{as}}\;\, x \to - \infty, \label{bc1}\\
\Phi (x) &\sim& e^{-i\omega x}\;\, {\mathrm{as}}\;\, x \to + \infty.\label{bc2}
\eea
We shall use Eqns.\ (\ref{scho}-\ref{qr}) and the boundary conditions
given above together with the metric given by Eqn.\ (\ref{metric}) to
obtain the highly damped asymptotic quasinormal modes for the $d$
dimensional Schwarzschild black hole with the Gauss-Bonnet correction.

Following ref. \cite{motl}, we consider the Eqn. (\ref{scho}) extended to
the whole complex plane. For small values of $r$, i.e. in the neighbourhood of 
$r=l$ where $l$ is arbitrarily small, the tortoise coordinate has the
form
\be 
x \sim \frac{r^{2d-3}}{4\alpha M^2 (2d-3)}. \label{tortoise}
\ee
In the same region, the leading singular term of the potential is of
the form
\bea
V(r[x])&=&\big[- 32+\frac{16}{(d - 3)^2} + \frac{48}{(d - 3)} + 40d -
\frac{32d}{(d - 3)^2}\nonumber\\
&& - \frac{96d}{(d - 3)} - 12d^2 + \frac{16d^2}{(d -
  3)^2}+\frac{48d^2}{(d - 3)}\big]\frac{1}{16x^2(2d -
  3)^2}\label{potl} 
\eea
Following \cite{motl}, the potential can be written as
\bea
V(r[x])=\frac{j^2-1}{4x^2} \nonumber
\eea
where $j=\frac{(d-1)(d^2+6d-23)^{\frac{1}{2}}}{(2d^2-9d+9)}$.  Let us
here note that for the Schwarzschild metric in $d$ space-time
dimensions, the value of $j = 0$ \cite{motl}, which is very different
from the Gauss-Bonnet case under consideration. We shall make further
remarks about this point later in the paper.

For highly damped asymptotic quasinormal modes, we take the frequency
$w$ to be approximately purely imaginary. Thus, for the Stokes line
defined by ${\rm Im}~ (wx) = 0$, we see that $x$ is approximately
purely imaginary. This together with Eqn. (\ref{tortoise}) implies
that for small $r$, we have $(4d-6)$ Stokes lines labeled by $n = 0,~ 1,~
2,\cdots , 4d - 7$. The signs of $(wx)$ on these lines are given by
$(-1)^{(n+1)}$ and near the origin, the Stokes lines are equispaced by
an angle $\frac{\pi}{2d-3}$. Also note that near infinity, $x \sim r$
and ${\rm Re}~(x) = 0$ and ${\rm Re}~(r) = 0 $ are approximately
parallel.  Thus two of the Stokes lines are unbounded and go to
infinity. In the limit of small $\alpha$, just as in the case of
Schwarzschild black hole, two more Stokes lines starting from the
origin would form a closed loop encircling the real horizon in the
complex $r$ plane \cite{natario}. The rest of the Stokes lines too
would have a structure qualitatively similar to those in the
Schwarzschild case.

We now proceed with the calculation of the asymptotic quasinormal
modes. The solution of the wave Eqn. (\ref{scho}) is given by
\be
\Phi(x) =  A \sqrt{2\pi\omega x}J_{\frac{j}{2}}(\omega x)+
B \sqrt{2\pi\omega x}J_{-\frac{j}{2}}(\omega x),\label{phi}
\ee
where $J_\nu$ is the Bessel function of first kind and $A,~B$ are
constants. Since we are considering the situation where ${\rm Im}~(w)
\rightarrow \infty$, we can use the asymptotic expansion of the Bessel
function to write the solution (\ref{phi}) as
\be
\Phi (x) = \left( A e^{-i\alpha_+} + 
B e^{-i\alpha_-}\right) e^{i \omega x} + 
\left( A e^{i\alpha_+} + B e^{i\alpha_-}\right) e^{-i \omega x}, \label{solution}
\ee
where $\alpha_\pm = \frac{\pi}4 (1 \pm j)$.  Now consider a point
$z_-$ near $r \sim \infty$ and situated on one of the unbounded Stokes
lines which is asymptotically parallel to the negative imaginary axis
in the complex $r$ plane. On such a point $z_-$, we have $wx
\rightarrow \infty$ and thus the asymptotic form of the solution
(\ref{solution}) is valid. Imposing the boundary condition (\ref{bc2})
we get from (\ref{solution}) that
\be
A e^{-i\alpha_+} + B e^{-i\alpha_-} = 0.\label{zero}
\ee
Consider now a point $z_+$ again near $r \sim \infty$ and situated on
one of the unbounded Stokes lines which is asymptotically parallel to
the positive imaginary axis in the complex $r$ plane. From the
geometry of the Stokes lines and their equispaced nature near the
origin, it is easy to see that in order to pass from the point $z_-$
to $z_+$ while always staying on the Stokes lines, it is necessary to
traverse an angle $\frac{3 \pi}{2d - 3}$ in the complex $r$ plane,
which amounts to a $3 \pi$ rotation in the tortoise coordinate $x$.
Using the analytic continuation formula for the Bessel function
\be
\sqrt{2\pi e^{3\pi i} \omega x}\ J_{\pm\frac{j}{2}} 
\left( e^{3\pi i} \omega x \right) = 
e^{\frac{3\pi i}2 (1 \pm j)}\sqrt{2\pi \omega x}\ J_{\pm\frac{j}{2}} 
\left( \omega x \right),
\ee
the solution at the point $z_+$ can be written as
\be
\Phi (x) = \left( A e^{7i\alpha_+} 
+ B e^{7i\alpha_-}\right) e^{i \omega x} 
+ \left( A e^{5i\alpha_+} + B e^{5i\alpha_-}\right) e^{-i \omega x}.
\ee
We now close the two asymptotic branches of the Stokes lines by a
contour along $r \sim \infty$ on which ${\rm Re}~(x) > 0$. Since we
are considering modes with ${\rm Im}~(w) \rightarrow \infty$, on this
part of the contour $e^{i w x}$ is exponentially small. Thus, for the
purpose of monodromy calculation, we rely only on the coefficient of
$e^{-i w x}$ \cite{motl}.  As the contour is completed this
coefficient picks up a multiplicative factor given by
\be
\frac{A e^{5i\alpha_+} + B e^{5i\alpha_-}}
{A e^{i\alpha_+} + B e^{i\alpha_-}}.
\ee
The monodromy of $e^{-i w x}$ along this clockwise contour is
$e^{-\frac{\pi\omega}{k}}$ where $k = \frac{1}{2} f^{\prime}(r_h)$ is
the surface gravity at the Gauss-Bonnet real horizon $r_h$. Thus the
complete monodromy of the solution to the wave equation along this
clockwise contour is
\be
\frac{A e^{5i\alpha_+} + B e^{5i\alpha_-}}{A e^{i\alpha_+} + 
B e^{i\alpha_-}} e^{-\frac{\pi\omega}{k}}. \label{mono}
\ee
The contour discussed above can now be smoothly deformed to a small
circle going clockwise around the horizon at $r = r_h$. Near $r = r_h$
the potential in the wave equation approximately vanishes. From the
boundary condition (\ref{bc1}), we see that the solution of the wave
equation (\ref{scho}) near the black hole event horizon is of the form
\be
\Phi(x) \sim C e^{i \omega x} \label{cond}
\ee
where $C$ is a constant. The monodromy of $\Phi$ going around the
small clockwise circle around the event horizon is thus given by
$e^{\frac{\pi\omega}{k}}$. Since the two contours are homotopic, the
monodromies around them are equal. Thus, from Eqn. (\ref{mono}) and
Eqn. (\ref{cond}) we
have
\be
\frac{A e^{5i\alpha_+} + B e^{5i\alpha_-}}{A e^{i\alpha_+} + B
  e^{i\alpha_-}} e^{-\frac{\pi\omega}{k}} =
e^{\frac{\pi\omega}{k}}.\label{final} 
\ee
Eliminating the constants $A$ and $B$ from Eqn. (\ref{zero}) and Eqn.
(\ref{final}) , we get
\be
{\rm e}^{\frac{2 \pi w}{k}} = -  \frac{{\rm sin}(\frac{3 \pi}{2})j}{{\rm
sin}(\frac{\pi}{2})j}.
\ee
Equivalently, we have,
\be
w = T_H{\rm log} \left | \frac{{\rm sin}(\frac{3 \pi}{2})j}{{\rm
sin}(\frac{\pi}{2})j} \right | 
+ 2 \pi i T_H \left ( n + \frac{1}{2} \right ), \label{freq}
\ee
where $T_H = \frac{k}{2 \pi} $ is the Hawking temperature of the black
hole (in units of $\hbar = 1$.) Eqn. (\ref{freq}) with
$j=\frac{(d-1)(d^2+6d-23)^{\frac{1}{2}}}{(2d^2-9d+9)}$ provides an
analytic expression for the highly damped asymptotic quasinormal modes
of the Gauss-Bonnet black hole in the limit where the Gauss-Bonnet
coupling constant $\alpha$ is small.  In the next Section, we discuss
some of the features and implications of the result obtained above.

\sxn{Discussion}

In this Paper we have calculated the asymptotic quasinormal
frequencies of the $d$ dimensional Schwarzschild black hole in
presence of a small Gauss-Bonnet correction term, where we have used
the string inspired assumption of the existence of a minimum fundamental 
length scale $l$. Our analysis is relevant only when the Gauss-Bonnet 
coupling $\alpha$ is suitably restricted such that ${8\alpha M}<<l^{d-1}$. 
For a massive black hole, this condition restricts $\alpha$ to very small values. This implies that the geometry that we are considering is very close to the Schwarzschild background with the Gauss-Bonnet term providing a small modification of the Schwarzschild geometry. 

The quasi-normal mode frequencies are functions of the spacetime dimension $d$ 
and they also depend on the metric parameters $M$ and $\alpha$ through the surface
gravity $k$. It may be noted that if the limit $\alpha \rightarrow 0$
is taken after the calculation of the asymptotic quasinormal mode,
then the resulting quasinormal frequencies do not tend towards those
for the Schwarzschild black hole for which $j = 0$. The reason for
this is not difficult to understand. The point is that in the
calculation of the asymptotic quasinormal mode, the short distance
singularity structure of the potential in (3.2) plays a crucial role
\cite{motl}. In presence of a small but nonzero value of the
Gauss-Bonnet coupling $\alpha$, the coefficient of the leading short
distance singularity is very different from that when $\alpha = 0$.
Thus, in the limit when $\alpha \rightarrow 0$, even though the
surface gravity for the Gauss-Bonnet black hole smoothly goes over to
that for the Schwarzschild case, the asymptotic quasinormal
frequencies do not.  The situation here is very similar to the well
known fact that the asymptotic quasinormal frequencies of the
Reissner-Nordstr\"{o}m black hole do not tend to those for the
Schwarzschild in the limit where the charge is taken to zero
\cite{motl,natario}. Thus, if the Schwarzschild black hole is
considered to be the zero charge limit of the Reissner-Nordstr\"{o}m
geometry, and if the Schwarzschild quasinormal frequencies are to be
obtained from analyzing the Reissner-Nordstr\"{o}m system, then the
charge equal to zero limit in the Reissner-Nordstr\"{o}m metric has to
be taken before the calculation of the quasinormal frequencies.
It may also be noted that approximations in the 
metric which change the nature of the singularity structure of the
differential equation have been used previously in the literature as
well for the purpose of making of quasinormal modes estimations
\cite{siopsis}.

In contrast to the Schwarzschild case, the real part of the asymptotic
quasinormal frequencies associated with the tensorial perturbations of
the Gauss-Bonnet black hole for small values of $\alpha$ do not
exhibit any universality with respect to the number of space-time
dimensions and is unlikely to have that feature for other type of
perturbations as well. However, as the number of dimensions is
increased, the real part of the quasinormal frequencies tend towards
zero. The conjecture of Hod \cite{hod} therefore does not seem to be
valid in presence of a small but finite Gauss-Bonnet coupling
$\alpha$. Our result thus provides a clear distinction between Einstein
gravity and its string derived variant.

The metric given by $f(r)$ in Eqn. (2.6) certainly does not capture the full effect 
of the general Gauss-Bonnet term for arbitrary values of the coupling $\alpha$. 
It would be nice to obtain an analytic expression for the asymptotic
quasinormal frequencies of the Gauss-Bonnet black hole for a general
value of the coupling $\alpha$, which has so far not been possible. It
would also be very important to obtain the quasinormal frequencies of
this system numerically.  The asymptotic quasinormal frequencies for
related string derived models of gravity, especially those with
cosmological constant and any associated connection to the dual
conformal field theories would also be interesting to find. Some of
this work is currently under progress \cite{fut}.

\section*{Acknowledgment}
The authors wish to  thank I. P. Neupane for useful comments.

\bibliographystyle{unsrt}

\end{document}